\documentclass[11pt]{article}
\usepackage{amssymb,amsmath}
\usepackage{psfig}
\usepackage{doublespace}

\begin{document}

\title{\bf MOTOR PROTEINS HAVE HIGHLY CORRELATED BROWNIAN ENGINES}
\author{G. P. Tsironis\\
Physics Department, University of Crete \\
and Foundation for Research and Technology -- Hellas, \\
P.O. Box 2208, 71003 Heraklion, Crete, Greece\\
Tel. +30-81-394220, FAX +30-81-394201, email gts@physics.uch.gr \\
\and
Katja Lindenberg \\
Department of Chemistry and Biochemistry 0340 \\
and Institute for Nonlinear Science, \\
University of California San Diego,\\
La Jolla, CA 92093-0340,USA\\
Tel. (619) 534-3285, FAX (619) 534-7244, email klindenberg@ucsd.edu
}

\date{\today}
\maketitle

\begin{abstract}
Two headed motor proteins, such as kinesin and  dynein,  hydrolyze
environmental ATP in order to propel unidirectionally along
cytoskeletal filaments such as microtubules.
In the intensely studied case of kinesin, protein heads of approximate 
dimension $4\times 4 \times 7~{\rm nm}^3$ bind primarily on the
$\alpha$ tubulin site of asymmetric $\alpha$--$\beta$ $8$nm-long tubulin
dimers that constitute the microtubular  protofilaments.  Kinesin
dimers overcome local binding forces up to approximately $5$pN and
are known to move on protofilaments with  ATP concentration-dependent
speeds in the range of $100-500$nm/sec while hydrolyzing on average
one ATP molecule per $8$nm step.  The salient features of protein
trajectories are the distinct abrupt  usually $8$nm-long steps from
one tubulin dimer to the next interlaced with long quiescent
binding periods at a tubilin site.  Discrete walks of this type
are characterized by substantially reduced variances compared to
pure biased random walks, and as a result rule out flashing-type
ratchet models as possible mechanisms for motor movement. On the other
hand, simple additive correlated brownian ratchets that present
exactly these discrete trajectory patterns with reduced variances
are compatible with the general features of the protein motion. 
In the simplest such model the protein is simplified to a single particle 
moving in a periodic non-symmetric tubulin-derived
potential and the environmental and ATP interaction is included in a 
correlated additive noise term.  For this model we show that
the resulting protein walk has features resembling experimental data.  
Furthermore, in more realistic mechanical models of two
masses connected by a spring we find qualitative agreement with
recent experimental facts related to motion of protein chimaeras
formed through kinesin motor domains with non-claret disjunctional
(ncd) neck regions.

\end{abstract}

\begin{center}
{\bf I. INTRODUCTION}
\end{center}

The attempt to understand the detailed mechanism of motor protein motion
in the cytoskeleton has led to the study of several stochastic ratchet
(references in our bibliography not explicitly cited are important
recent experimental and modeling contributions).
The simplest such model
involves one overdamped particle representing, for instance, the 
motor protein kinesin, moving in a periodic but not symmetric force
field, driven by different types of correlated
noises [Magnasco 1993, 1994; Astumian and Bier; Astumian;
Doering et al.; Millonas and Dykman;
Bartussek et al.].  The periodic forces
are exerted by the $8$nm-long $\alpha$--$\beta$ partially
asymmetric  tubulin dimers on kinesin while the noise terms represent
the fluctuating environment. This model leads to a macroscopic
particle current in a specific direction determined by the
potential asymmetry and by the properties of the noises. 
In the colored noise case, the finite
correlation time $\tau$ corresponds to the ATP kinesin binding event time
and subsequent energy release through hydrolysis. 
The ATP hydrolyis rate is approximately $50 {\rm s}^{-1}$ and results
on average in one $8$nm kinesin step per ATP cycle, i.e. to a typical
kinesin speed of $400$nm/s [Schnitzer and Block; Hua et al.].
The single-particle colored noise model in the highly correlated
noise regime leads to a natural interpretation for the 
kinesin steps observed in experiments.  In this regime the 
brownian particle simply waits in the potential minimum of a 
tubulin unit until the appropriate fluctuation arrives that allows
it to escape to the next tubulin dimer
[Tsironis and Grigolini; de la Rubia et al.; Lindenberg et al.; 
Dykman and Lindenberg].
The noisy environment is the ambient fluid containing a variety of
molecules, inlcuding ATP molecules at $\mu$M concentration levels. 
A large number of unsuccessful binding attemps of ATP molecules contribute
to medium fluctuations, while the successful critical noise fluctuation can
be interepreted as an ATP successfully binding on kinesin.  The
critical binding fluctuation determines the average
exit time $\left<T\right>$, or average step time, leading
to a distance $x$ after $n$ steps, i.e.  $x \approx n \left<T\right>$.
\\
\\

\centerline{
\hspace*{1.2cm}
\hbox{
\hspace*{1.2cm}
\psfig{file=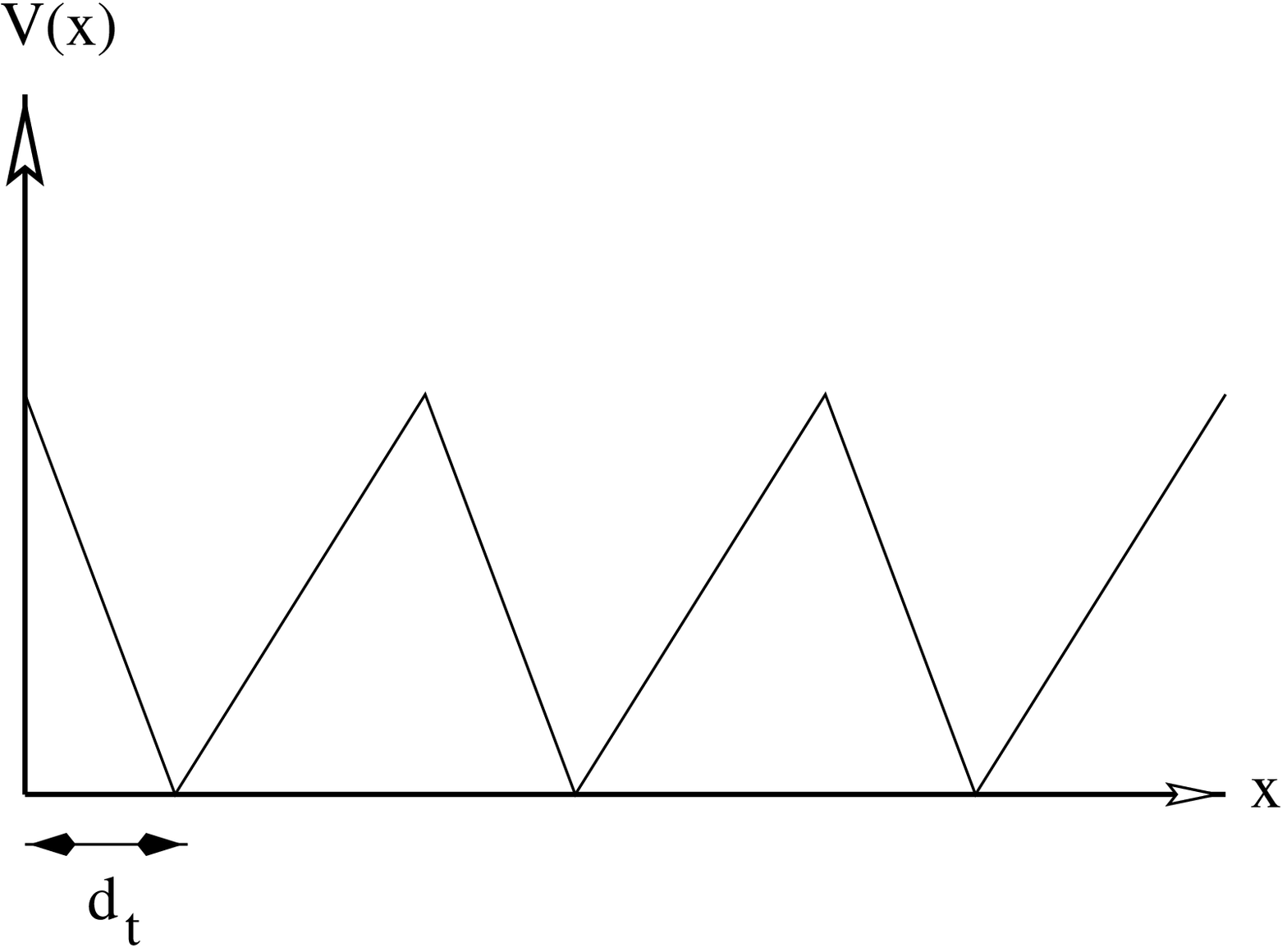,height=7.0cm,width=9.0cm,angle=270}
}}

\begin{singlespace}
\noindent
{\bf Figure 1}
Asymmetric periodic potential used for analysis.  The potential is
piecewise linear and of height $V_0$.  The distance between the peak
and the minimum along the steeper portion of the potential is $d_t$.
\end{singlespace}

\clearpage

\begin{center}
{\bf II. HIGHLY CORRELATED BROWNIAN MOTOR PROTEIN VELOCITY ESTIMATES}
\end{center}

The calculation of the average step time $\left< T\right>$
proceeds as follows.  The equations for the overdamped kinesin motion are
\begin{eqnarray}
\gamma  \frac{dx}{dt} &=& f(x) + \xi(t), \nonumber   \\  &&\nonumber\\
 \frac{d \xi}{dt} &=& - \frac{1}{\tau} \xi + \frac{1}{\tau} \eta (t),
\label{eq1}
\end{eqnarray}
where $x(t)$ is the kinesin position and
$f(x) \equiv - V^{\prime}(x)$ where $V(x)$ is
the periodic non-symmetric potential of the protofilament.  
We use the simple piecewise linear potential shown in Figure~1.
The height of the potential is $V_0$ and the distance from bottom to
top on the steep side is denoted $d_t$. The auxiliary variable $\xi$
represents the coupling of the particle to the environment; if the
noise variable $\eta (t)$ is
Gaussian and delta-correlated, $\left<\eta (t) \eta (t')\right> =
 D \gamma \delta (t-t')$, then $\xi(t)$ is an Ornstein-Uhlenbeck
process, that is, $\xi(t)$ is Gaussian and exponentially correlated,
\begin{equation}
 \left<\xi (t) \xi (t') \right> = \frac{D \gamma }{2 \tau}
e^{- \frac{|t-t'|}{\tau}}.
\label{2eq2}
\end{equation}
The white noise strength $D$ is taken to be the ambient temperature
multiplied by the Boltzmann constant, i.e., $D=k_B T$,
and $\gamma$ is the medium damping. Typical trajectories
resulting from Eq.~(\ref{eq1})
are shown in Figure~2 for different, but all large, noise correlation
times.  There is a dramatic
difference in the stochastic motor trajectory for small
(white noise limit) and large 
(deterministic limit) correlation times $\tau$.
A short correlation time reflects a large number of successful
ATP binding events that propel the stochastic motor onward. 
Indeed, with short correlation times the tubulin site motor
residence time and the jump time are comparable, leading to a
unidirectional (due to the high asymmetry in the tubulin binding
force) fluctuating non-step-like trajectory.
At high correlation times, on the other hand, the nature of the
walk is quite different, characterized
by long waiting periods at a site until the appropriate
critical ATP binding fluctuation
arrives and enables the motor to perform a step that is quite abrupt
relative to the waiting time.  The waiting time or exit time
$\left<T\right>$ for a successful fluctuation to occur increases with the
degree of the temporal correlation of the noise (reflecting slower ambient
changes). This time thus determines the average clock rate for the motion.

\centerline{
\hspace*{1.2cm}
\hbox{
\hspace*{1.2cm}
\psfig{file=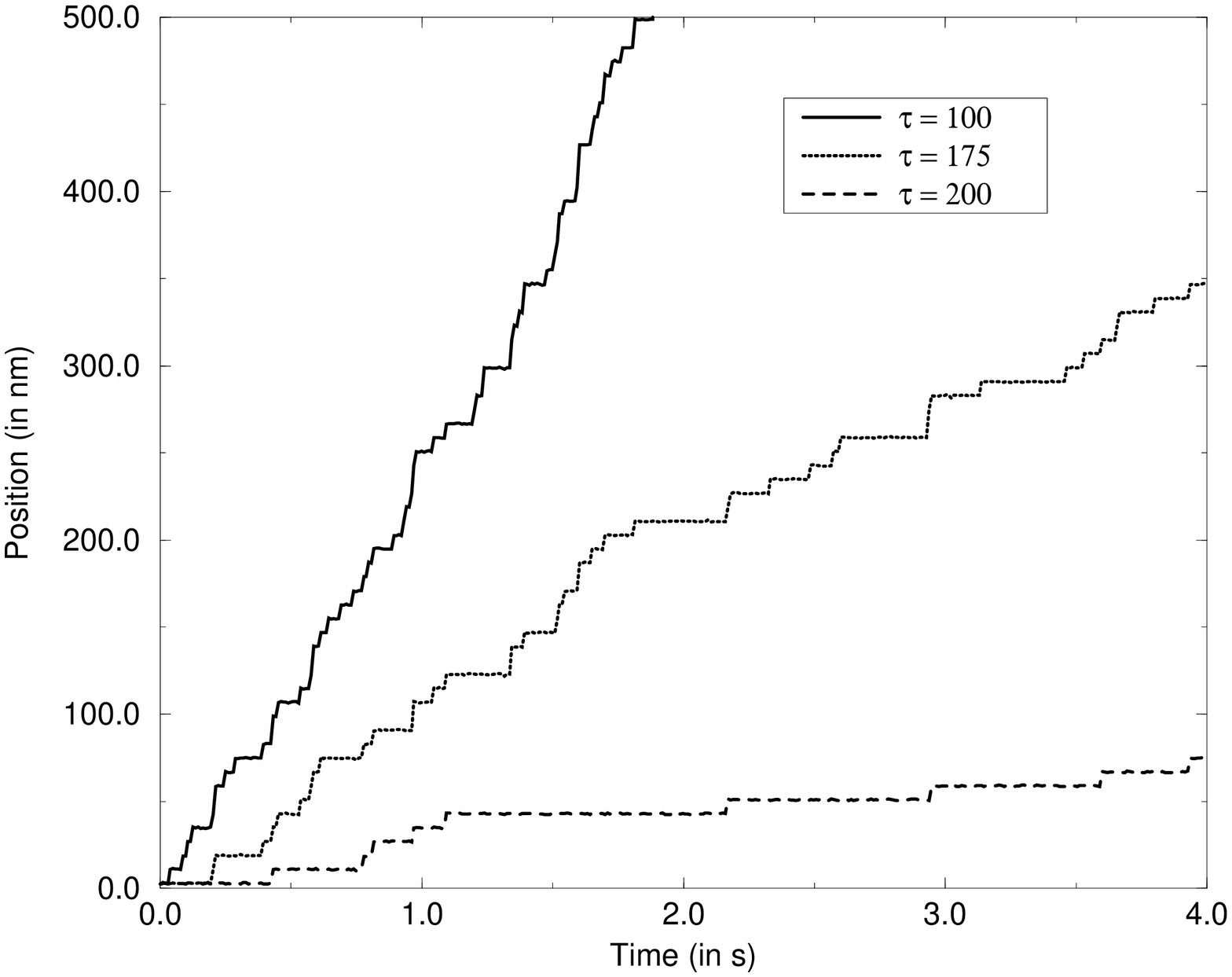,height=7.0cm,width=9.0cm,angle=270}
}}

\begin{singlespace}
\noindent
{\bf Figure 2}
Position traces (in nanometers) vs time (in seconds)
for highly correlated, quasi-deterministic
brownian engines showing the characteristic step-like forward motion
of molecular motor proteins.  The noise correlation time $\tau$ (in
nanoseconds) is the unique adjustable parameter of the model.
Relatively large correlation times are clearly associated with a
quasi-deterministic step-like process, more markedly so with increasing
correlation time.  For these trajectories we used a potential barrier
$V_0 = 10 k_B T$ and $d_t = 2.4$nm, leading to forces of approximately
$16.6$pN and $-7.1$pN.  As the correlation time decreases, the
steps become shorter, and the motion becomes faster but locally
more erratic.
\end{singlespace}


To find a rough analytical estimate for the mean exit time from one
tubulin dimer to the next that serves to illustrate the parameter
dependences, we assume that kinesin binds primarily to the
$\alpha$-tubulin, and that the relatively large asymmetry of the
hypercell tubulin potential justifies ignoring the
exit times toward the high force  direction because of their relatively
high improbability.  The mean exit time along the forward (lower force)
direction then is
\begin{equation}
\left< T \right> ~=~ \frac{\sqrt{2 \pi D \gamma \tau}}{|{\xi_c} |}~ \exp
\left( \frac{{{\xi^c}}^2 \tau}
 { 2 D \gamma }\right),
\label{eq2}
\end{equation}
which is just the so-called Kramers time for the fluctuating force
$\xi$ to reach the critical value $\xi_c$ for the first time.
The critical force $\xi_c$ necessary for cancelling the
tubulin force $f(x)$ and rendering kinesin free to move
to the next tubulin hypercell is equal to
$\xi_c = V_0/d_t+\gamma d_t /\tau$.  The first ``dominant" 
term is the total force needed to reach the potential
maximum $V_0$ while the second
``correction" term  permits kinesin to move to the next cell before the
highly correlated force $\xi$ aqcuires a new value and most likely
interrupts its exit flight.

The point we stress here is that the average residence time
depends critically on the correlation 
time $\tau$  as its only adjustable parameter.  In our estimate
this dependence appears in the form 
$\left<T\right> = A( \tau ) \exp [ S( \tau )]$. 
With a potential barrier 
$V_0 \approx 8  k_B T$ and $d_t \approx 5.3$nm we have
$V_0 / d_t \approx 6$pN. These are realistic parameter values. With
the further typical values 
$D = k_B T \approx 4 \times 10^{-21}$J at room temperature and   
$\gamma = 6 \times 10^{-11}$kg/s we obtain $\xi_c \approx 6 {\rm pN}
 + 318 {\rm pN} /\tau [{\rm ns}]$,
where $\tau$[ns] is the correlation time in nanoseconds.
The exponent and prefactor are
\begin{equation}
S( \tau ) = \frac{{{\xi^c}}^2 \tau} { 2 D \gamma } = \frac{V_0}{D} + 
\left(\frac{V_0}{d_t}\right)^2 \frac{\tau}{2 D \gamma } + \frac{ \gamma {d_t}^2 }{2 D}
\frac{1}{\tau} 
~=~  8 + 0.075 \tau [{\rm ns}] + \frac{210}{\tau [{\rm ns}]}
\end{equation}
and 
\begin{equation}
A(\tau ) ~=~  \frac{\sqrt{2 \pi D \gamma \tau}}{|{\xi^c} |} ~\approx~
 6.5 {\rm ns}  \sqrt{ \tau [{\rm ns}]} 
\end{equation}
where in the prefactor we used only the dominant force term.
Finally
\begin{equation}
\left<T\right>
~=~ 6.5 ~{\rm ns}~  \sqrt{ \tau [ {\rm ns}]} ~
e^{8+ 0.075 \tau [{\rm ns}]+{210}{\tau} } .
\end{equation}
We specifically note the dramatic exponential dependence of the average
step time on the
correlation time $\tau $. For these numbers a correlation
time of $\tau \approx 100$ns
leads to the average exit time $\left<T\right> \approx 2$s
or a kinesin speed of $4$nm/s.  Although subtantially smaller than
observed values due to the asymptotic character of our approximation,
it is nevertheless an informative estimate.  Indeed, 
numerical simulations for similar parameter ranges give
smaller mean exit times, leading to kinesin speeds one to two orders of
magnitude larger, results that are compatible with
experimental data.  For example, in Figure~2, which corresponds to an even
higher barrier, one obtains a kinesin speed of $\sim 200$nm/s for
$\tau=100$ns.  

Due to its quasi-deterministic character,  
the simple brownian ratchet model for kinesin in the high
correlation regime leads to a
reduced position variance as opposed to flashing ratchet models. 
In the latter,
the source of the nonequilibrium fluctuations is the stochastic
on--off switching
of the entire ratchet potential (multiplicative fluctuations). 
In that case, the brownian particle
remains in the vicinity of the binding site while the potential is on,
and moves diffusively when the potential temporarily disappears. 
As a result, the position trace is
characterized by a variance closer to that of a free diffusion process 
that involves not only forward steps but also 
multiple backward steps, a picture not
compatible with experimental data.  In the additive fluctuation
model, on the other hand,
the variance becomes substantially reduced with increasing correlation
time [Lindenberg and Tsironis], while backward slips are
extermely improbable even for small potential asymmetries. 

From the compatibility of the simple correlated brownian ratchet model with 
the qualitative and quantitative features of kinesin motion, a simplified
picture emerges for the movement of motor proteins on microtubles.  We find
that the main features of this picture are the asymmetric periodic
tubulin potential and the coupling to a stochastic environment that is
necessarily highly correlated and can even be said to be
quasi-deterministic.  The single adjustable parameter of the model 
is the noise correlation time that is physically related to the event times
for ATP binding at the active kinesin head as well as to the hydrolysis time.
For reasonable kinesin parameters this time is of the order of
$100$ns to $1\mu$s,
a reasonable range for the phenomena involved.  It is not the aim of
this simple brownian model to account fully for the specifics of the kinesin
walk by taking into account higher dimensional features.  It is,
however entirely possible to extend the model in this direction for 
more complete quantitative agreement with the data.

\begin{center}
{\bf III. MOTION OF MOTOR CHIMAERAS}
\end{center}

Kinesin and non-claret disjunctional (ncd) are kinesin
superfamily molecular motor proteins that move towards the plus
and minus ends of microtubules
respectively [Svoboda and Block; Svoboda et al. 1993, 1994; Howard]. 
Recent experiments with synthetic protein chimaeras show that while
the protein catalytic domain seems to be responsible for the processivity of
the motor on the microtubule, the ``neck" region adjacent to the motor heads  
controls the {\em directionality} of movement
[Case; Henningsen and Schliwa].
The simple one-particle model presented above is clearly not adequate to
describe these experimental facts
either quantitatively or even qualitatively.  Although
some attempts in the direction of stochastic modeling of
protein motion reversals have been made,
the high degree of correlations in the ATP fueling process make it necessary
to consider quasi-deterministic models.  We argue that
a simple newtonian model of two-motor-head particles connected
through a neck coiled-coil spring whose rest length changes
with each ATP hydrolysis event
captures the essential motor dynamics features.  In particular, the observed 
directionality reversal in chimaeras with different coiled-coil regions
appears
in the model from a change in the stiffness of the spring constant.  
Motor speed is determined by the average ATP absorption rate while the
effect of ambient temperature is very small, leading to an
essentially non-brownian deterministic motor [Dialynas et al.].
\\

\centerline{
\hspace*{1.2cm}
\hbox{
\hspace*{1.2cm}
\psfig{file=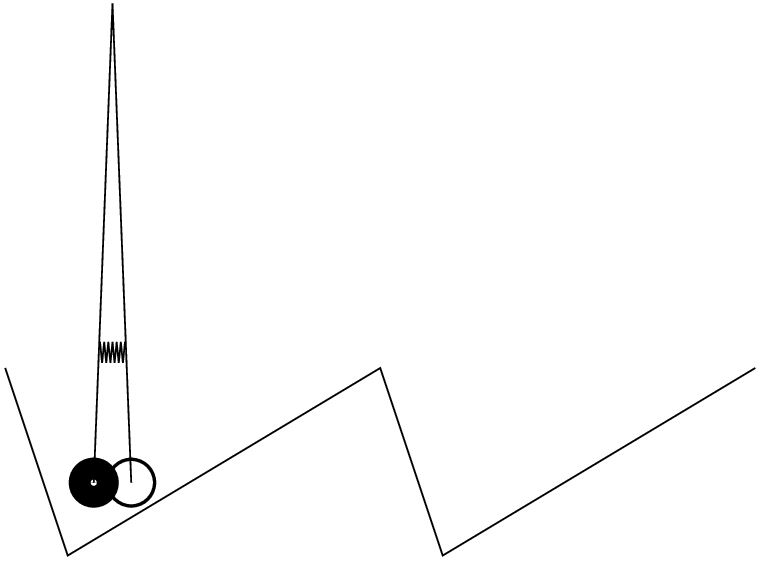,height=3.0cm,width=3.0cm,angle=270}
\hspace*{0.4cm}
\psfig{file=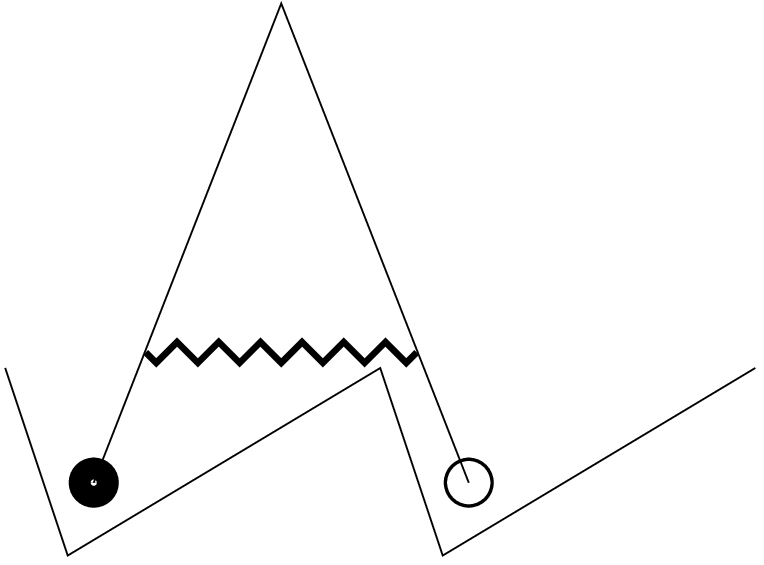,height=3.0cm,width=3.0cm,angle=270}
\hspace*{0.4cm}
\psfig{file=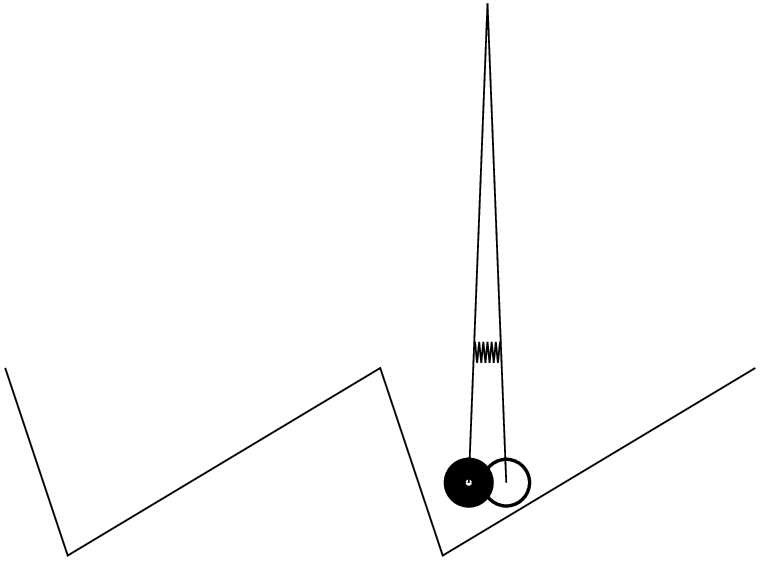,height=3.0cm,width=3.0cm,angle=270}
}
\hspace*{1.5cm}}
\hspace*{.7cm}
\hspace*{3.05cm}
{\bf a} \hspace*{3.05cm}
{\bf b} \hspace*{3.05cm}
{\bf c}

\begin{singlespace}
\noindent
{\bf Figure 3}
Deterministic motor dimer walk associated with one ATP per step when the
coiled-coil spring is relatively soft (see text). The motor moves
towards the easier macrotubule potential slope as described in this
sequence of figures. Open circle denotes the ``right" head.
(a) The dimer is in the relaxed state associated
with a shrunken spring.  Both heads are
in the same well. (b) An ATP hydrolysis event causes the coiled-coil spring
to open.  This causes the ``right" head to move toward the right.  The
relaxed state has the heads in adjacent wells. (c) ADP release leads to the
closure of the spring, which in turn causes the ``left" head to rejoin the
``right" one in a single well. The entire dimer has thus moved to the
right and is ready for a repetition of the cycle.
\end{singlespace}

The  model we consider is a deterministic version of the stochastic
model of Der\'enyi and Vicsek.  We show qualitatively that
it can produce newtonian motion of proteins in different
directions [Stratopoulos et al.].
In the model  $x_1$, $x_2$ denote the positions of the two dimer
heads interacting with
the microtubule surface through a one dimensional periodic
non-symmetric potential with
a unit cell of length $8$nm.  The two dimer heads also interact
with each other through an internal harmonic potential 
$\; V_s(x_1,x_2) \;=\; \tfrac{1}{2} \kappa (|x_1 - x_2| - l(t))^2 $. Here
$l(t) \,$ is the time-dependent rest length of the spring  
that takes on two values corresponding to 
two different states of the dimer. In state (a)\, $l(t) = l_1 = 0 $,
that is, the two heads are close to each other. After an 
ATP hydrolysis event, a protein conformational change takes place and 
the dimer transits to state (b), characterized by an opened $\alpha$-helical
coiled-coil of length $\, l(t) = l_2 \, = 8$nm. 
Once ADP is released, the dimer transits back
to state (a) and the conformational change cycle repeats,
with an ATP hydrolysis
rate taken to be approximately $\, 50{\rm s}^{-1}$.  
The assumption of a deterministic periodic $l(t)$ versus a
stochastic dichotomous time dependence is not important and
leads to qualitatively similar results. 
We thus consider that the dimer spends equal times $t_a=t+b$ in 
each state leading to a period of the conformational change cycle
$\, t_c = t_a+t_b =0.02$s.

Let us first examine the case when the spring connecting the protein heads
is soft. Initially 
$l(t) = l_1 = 0$ and the two heads relax to the bottom of the left well as 
shown in Figure~3a. When ATP binds and the coiled-coil unwinds, the
rest length of the spring changes to 
$l(t) = l_2 $ while the spring tension tends to move the two heads apart 
in order to relax the spring. For an ultra soft spring, 
the spring tension $f_{spring}$ cannot surmount the microtubule
potential forces $f^r$ (caused by the steeper side of the potential, and
hence the stronger of the two) and $f^l$, and both heads remain
trapped at the bottom of the well.
For stiffer springs with tension force in the range between the two constant
force values of the piece-wise linear microtubule potential,
$f^l < f_{spring} < f^r$,
the left head tries to move to the left  due to the spring
force but the opposing microtubule force $\; f^r$ 
is too strong.  The left head thus remains at the bottom of 
the well. On the other hand, the spring pushes the right head toward
the right and since its tension is stronger than the potential
force, the right head moves to the right  until the spring relaxes 
(Figure~3b). At that stage the distance between the two heads 
is exactly $l_2$, i.e., equal to the period of the potential.
The spring is now relaxed (state (b)), while at the same time
each single head relaxes to the bottom of two adjoining wells.
Note that implicit in this analysis
is the assumption that the spring remains in its extended
state for a sufficiently
long time for the right head to reach the neighboring well.
Subsequently the motor spring again makes a transition to the
shrunken state and thus 
the two heads again tend to move closer to each other. Again,
the right head cannot move to the left, while the left head
moves to the right
until it reaches the right head (Figure~3c). At this point
the spring is relaxed to its shrunken state (state (a)), and at the
same time both heads relax at the bottom of the right well.  As a result
of this purely mechanical cycle the motor protein has consumed one ATP molecule
and has advanced one step to the right.
\\

\centerline{
\hspace*{1.2cm}
\hbox{
\hspace*{1.2cm}
\psfig{file=gr1.eps,height=3.0cm,width=3.0cm,angle=270}
\hspace*{0.2cm}
\hspace*{0.2cm}
\psfig{file=gr3.eps,height=3.0cm,width=3.0cm,angle=270}
\hspace*{0.2cm}
\hspace*{0.2cm}
\psfig{file=gr5.eps,height=3.0cm,width=3.0cm,angle=270}
}
\hspace*{1.5cm}}
\hspace*{.7cm}
\hspace*{3.05cm}
{\bf a} \hspace*{3.05cm}
{\bf b} \hspace*{3.05cm}
{\bf c} 

\begin{singlespace}
\noindent
{\bf Figure 4}
Deterministic motor dimer walk associated with one ATP per step when the
coiled-coil spring is hard (see text). The motor moves
towards the steeper macrotubule potential slope as described in this
sequence of figures. Open circle denotes the ``right" head.
(a) The dimer is in the relaxed state associated
with a shrunken spring.  Both heads are
in the same well. (b) An ATP hydrolysis event causes the coiled-coil spring
to open.  This causes the ``left" head to move toward the left.  The
relaxed state has the heads in adjacent wells. (c) ADP release leads to the
closure of the spring, which in turn causes the ``right" head to rejoin
the ``left" one in a single well. The entire dimer has thus moved to
the left and is ready for a repetition of the cycle.
\end{singlespace}

For the strong spring case, $\; f_{spring} \gg \; f^r > \; f^l$,
we consider again
the two heads initially relaxed but in the right well 
as in Figure~4a, i.e., $l(t) = l_1$. When the spring transits from state (a) to 
state (b), the spring tension pushes the two heads apart. Since the potential 
forces are small compared to the spring tension, the right and the left 
heads move to the right and left respectively with almost equal speed
until the spring is relaxed. Because of the asymmetry of the potential,
at the time the spring is relaxed in its extended state
the left head has already jumped to the left neighboring well while the
right head is still in the original well. The 
spring is now relaxed and the two heads roll to the bottom of the
wells until
they relax at the bottom of the respective wells (Figure~4b). 
When now the spring transits back to the shrunken state, the two
heads move towards each other with almost equal speed. again due to the
asymmetry, at the time that the two heads meet (i.e., when the
spring has relaxed), the right head has already jumped to the adjacent
left neighboring well while the the left head remains in that same
(left) well. Finally the two heads relax to the bottom of the left
well (Figure~4c).  In this case
after one ATP cycle the motor protein has moved one step to the left.

This qualitative analysis, supported by a quantitative analysis
of this quasi-determinsitc model [Stratopoulos et al.],
indicates that the specific value of the spring constant connecting the
two protein dimers plays a decisive role in the selection of
directionality of motion.
This is compatible with the protein chimaera experiments if we interpret
the latter in the following way:  Both kinesin and ncd are
characterized by an $\alpha$-helix
coiled-coil that can be modeled by linear springs each
of {\it different} spring constants.
This feature enables them to move in oposite directions along
the {\it same} microtubule potential.  When a chimaera is
constructed, the protein neck region of kinesin, and therefore
its spring, is replaced by that of ncd while leaving the motor region
intact.  The model thus predicts motion reversal, a fact that is
readily supported by experiments.

\begin{center}
{\bf IV. CONCLUSIONS}
\end{center}

A simple single-particle stochastic ratchet model for motor proteins
shows reasonably good qualitative and even quantitative agreement
with biological experiments when
the correlation time of the noise is high.  In this regime,
the particle motion is controlled by a Kramers rate that depends strongly
(exponentially) on the correlations of the fluctuating environment. 
As a result, the protein motion
is essentially step-like and unidirectional, characterised by long
``waiting times" interrupted by fast steps. 
These motion features are rather deterministic, even though they
stem from a fundamentally fluctuating process. The results are compatible
with the experimental observations of reduced motion variances and
step-like protein motion.  The basic model of a motor protein moving in a 
very organized, quasi-deterministic way in the stochastic environment
of the cell is carried one step further with a more realistic
newtonian two-particle-and-spring model.  The two motor heads are each
in contact with 
the microtubule through a periodic non-symmetric potential, while the 
$\alpha$-helical coiled-coil interaction is modeled through a spring of
variable (binary) rest length.  The ATP hydrolysis process powers the engine
through a cycle that first unwinds the coil and subsequently rewinds it. 
As a result, the protein moves one step per ATP cycle in a direction that
depends on the value of the $\alpha$-helical spring constant.   A
kinesin-ncd chimaera is simply effected in the model when the 
$\alpha$-helix spring of one is replaced by that of the other, resulting,
as in the experiments, in direction reversal.  
Stochastic fluctuations do not affect the movement process substantially, 
thus reconfirming the basically deterministic nature of the process.    

\begin{center}
{\bf Acknowledgements}
\end{center}

We thank Larry Goldstein, W. E. Woerner
and Aris Moustakas for
helpfull discussions and G. Stratopoulos for computer assistance.
This work was supported in part by a $\Pi E N E \Delta $ 
Grant of the Greek Secreteriat for Research of the Ministry of
Development of Greece (G. P. T.), by the U. S. Department of Energy through
Grant No. DE-FG03-86ER13606 (K. L.),
and by NATO through Travel Grant 950399.

\begin{center}
{\bf REFERENCES}
\end{center}

\begin{description}

\item
Astumian, R. D. {\em Science} {\bf 1997},  {\em 276}, 917. 

\item
Astumian, R. D.; Bier, M.
{\em Phys. Rev. Lett.} {\bf 1994}, {\em 72}, 1766. 

\item
Bartussek R.; Reimann, P;  H\"{a}nggi, P.
{\em Phys. Rev. Lett.} {\bf 1996}, {\em 76}, 7.

\item
Case, R. B.; Pierce, D. W.; Hom-Booher, N.; Hart, C. L.; Vale, R. D.
{\em Cell} {\bf 1997}, {\em 90}, 959.



\item
Der\'enyi, I;  Vicsek, T.
{\em Proc. Natl. Acad. Sci. USA} {\bf 1996}, {\em 93}, 6775.

\item
Dialynas, T. E.;  Lindenberg, K.;  Tsironis, G. P.
{\em Phys. Rev. E} {\bf 1997}, {\em 56}, 3976.


\item
Doering, C. R.;  Horsthemke, W.;  Riordan, J.
{\em Phys. Rev. Lett.} {\bf 1994}, {\em 72}, 2984. 


\item
Dykman, M. I.; Lindenberg, K. in {\em Some Problems in Statistical
Physics}; Weiss, G. H., Ed.; SIAM: Philadelphia, 1993, p. 1.

\item
Henningsen, U.; Schliwa, M.
{\em Nature} {\bf 1997}, {\em 389}, 93.

\item
Howard, J.
{\em Ann. Rev. Physiol.} {\bf 1996}, {\em 58}, 703.

\item
Hua, W.; Young, E. C.; Fleming, M. L.; Geiles, J.
{\em Nature} {\bf 1997}, {\em 388}, 390.

\item
Lindenberg, K;  West, B. J.;  Tsironis, G. P. {\em Rev. Solid State
Science} {\bf 1989}, {\em 3}, 134. 

\item
Lindenberg, K.; Tsironis, G. P. Unpublished.

\item
Magnasco, M. O.
{\em Phys. Rev. Lett.} {\bf 1993}, {\em 71}, 1477.

\item
Magnasco, M. O.
{\em Phys. Rev. Lett.} {\bf 1994}, {\em 72}, 2656.

\item
Millonas, M. M.; Dykman, M. I.
{\em Phys. Lett. A} {\bf 1994}, {\em 185}, 65. 



\item
Schnitzer, M. J.; S. M. Block, S. M.
{\em Nature} {\bf 1997}, {\em 388}, 386.


\item
Stratopoulos, G. N.; Dialynas, T. E.; Tsironis, G. P. 
University of Crete preprint, unpublished.

\item
Svoboda, K.; Block, S. M.
{\em Cell} {\bf 1994}, {\em 77}, 773.

\item
Svoboda, K.; Mitra, P. P.; Block, S. M.
{\em Proc. Natl. Acad. Sci. USA} {\bf 1994}, {\em 91}, 11782.

\item
Svoboda K.; Schmidt C. F.; Schnapp B. J.; Block S. M.
{\em Nature} {\bf 1993}, {\em 365}, 721.



\item
Tsironis, G. P.; Grigolini, P.
{\em Phys. Rev. A} {\bf 1988}, {\em 38}, 3749.




\end{description}

\end{document}